\documentclass[pre,aps,twocolumn,superscriptaddress,showpacs]{revtex4}
\usepackage{amssymb}
\usepackage{epsfig}
\usepackage{amsmath}
\usepackage{subfigure}
\usepackage{graphicx}
\usepackage{textcomp}
\usepackage{url}
\usepackage{float}
\usepackage{color}

\begin{document}
\title{Mortality, Redundancy, and Diversity in Stochastic Search}

\author{Baruch Meerson} \affiliation{Racah Institute of Physics, Hebrew
  University of Jerusalem, Jerusalem 91904, Israel}

\author{S. Redner} \affiliation{Department of Physics, Boston University, Boston,
  MA 02215, USA and Santa Fe Institute, 1399 Hyde Park Road, Santa Fe, New
  Mexico 87501, USA}


\begin{abstract}

  We investigate a stochastic search process in one dimension under the
  competing roles of mortality, redundancy, and diversity of the searchers.
  This picture represents a toy model for the fertilization of an oocyte by
  sperm.  A population of $N$ independent and mortal diffusing searchers all
  start at $x=L$ and attempt to reach the target at $x=0$.  When mortality is
  irrelevant, the search time scales as $\tau_D/\ln N$ for $\ln N\gg 1$,
  where $\tau_D\sim L^2/D$ is the diffusive time scale.  Conversely, when the
  mortality rate $\mu$ of the searchers is sufficiently large, the search
  time scales as $\sqrt{\tau_D/\mu}$, independent of $N$.  When searchers
  have distinct and high mortalities, a subpopulation with a non-trivial
  optimal diffusivity are most likely to reach the target.  We also discuss
  the effect of chemotaxis on the search time and its fluctuations.

\end{abstract}

\maketitle

Stochastic searching~\cite{BLMV11} underlies a wide variety of processes in
biology~\cite{BWV81,V07,M08}, animal
foraging~\cite{C76,B91,OBE90,VBH99,BLMV06}, chemical
reactions~\cite{S17,HTB90}, and search operations for missing persons or lost
items~\cite{RS71,FS01,S09}.  A basic goal is to minimize the time needed to
successfully find a desired target.  Through evolution and algorithmic
developments, Nature and man have developed clever strategies to optimize
searches.  Many of these processes involve a single searcher that continues
the search for as long as necessary to reach the target.  In this Letter, we
investigate the role of fundamental influences on stochastic search that have
not yet received sufficient attention: mortality, redundancy, and diversity.
The notion that a searcher dies if it does not reach a target within a fixed
time epitomizes unsuccessful foraging.  While the properties of random walks
that die at a fixed rate has been recently
investigated~\cite{BNHW87,YAL13,AYL13,M15}, the general problem of how to
best conduct a search with mortal searchers is barely explored~\cite{BR14}.

When a searcher can die before reaching a target, a natural way to promote
success is by launching many searchers.  This redundancy embodies the
fertilization of an oocyte, where (in humans) roughly $3\times 10^8$ sperm
cells initially attempt to reach the oocyte after copulation.  Fertilization
is an extremely complex and multi-faceted process (see
Refs.~\cite{Eisenbach,Holcman} for reviews).  We make no pretense of
accounting for the many steps that a sperm undergoes to reach and
ultimately fertilize the oocyte.  However, sperm mortality, redundancy, and
diversity all play roles in this fertilization process.  Inspired by these
basic facts, we investigate an idealized scenario of fertilization that are
driven by the above attributes.

The geometry of the system is quite simple.  An oocyte located at the origin,
$x=0$, is represented by an absorbing boundary condition.  At time $t=0$, $N$
diffusing searchers, each of which represents a sperm cell, are launched from
$x=L$.  The basic questions that we address are: (i) What is the probability
of fertilization---that at least one of the searchers reaches the origin?  (ii) What
is the average time to reach the origin as a function of $N$ and the
mortality rate of the sperm?  If \emph{immortal} searchers are uniformly
distributed on the line, the survival probability of the target in one
dimension is known to decay as
$\exp(-\mathrm{const.}\times\sqrt{t})$~\cite{ZKB83,RK84,BZK84,Blythe,BMS13,RM,MVK}.
However, the biologically relevant situation where the searchers all start at
the same point has been less extensively investigated even when the searchers
are immortal \cite{Lindenberg80,RK99,KMR,Monasterio,MR}.

It is worth noting two important points: First, if the searchers are
immortal, then a basic result of first-passage processes is that a single
searcher will eventually reach the target, but the average time for this
successful search is infinite~\cite{R01}.  The average search time is again
infinite for $N=2$ searchers, but is \emph{finite} for $N\geq 3$
\cite{Lindenberg80}.  Second, for mortal searchers that die at a fixed rate,
the average time for a successful search, even for a single searcher, is
finite and approaches zero as the mortality rate increases.  In this
high-mortality limit, the only way for a searcher to reach the target is to
do so very quickly.

We first determine how a single mortal diffusing searcher finds the target.
Let $\rho(x,t)$ be the probability density that the searcher is located at
$0<x<\infty$ at time $t$.  Its dynamics is determined by the diffusion-decay
equation $\partial_t \rho =D\partial_x^2 \rho-\mu \rho$, where we assume a
fixed diffusivity $D$ and mortality rate $\mu$.  For a target at $x=0$ and a
searcher starting at $x=L$, the image method~\cite{R01} gives the solution to
this diffusion-decay equation as:
\begin{align}
\label{sol}
 \rho(x,t)=\frac{e^{-\mu t}}{\sqrt{4\pi Dt}}\, \left[e^{-(x-L)^2/4Dt}-e^{-(x+L)^2/4Dt}\right].
\end{align}
The flux at $x=0$ yields the probability density that the target
is reached at time $t$:
\begin{equation}
\label{FPP}
  f_1(t)=D \,\partial_x \rho(x,t) |_{x=0} =
  \frac{L}{\sqrt{4\pi D t^3}}\,\,  e^{-\mu t -{L^2}/{4 D t}}\,,
\end{equation}
while the probability $F_1(t) = \int_0^t dt^{\prime}\, f_1(t^{\prime})$ that
the searcher reaches the target by time $t$ is
\begin{align}
\label{exit-t}
  F_1(t) &= \tfrac{1}{2} \,e^{\sqrt{\frac{\mu}{D}}\,L} \mathrm{erfc}\big(L/\sqrt{4Dt}+\sqrt{\mu
  t}\big)\nonumber \\
  &~~~+\tfrac{1}{2} \,e^{-\sqrt{\frac{\mu}{D}}\,L}\,\mathrm{erfc}\big(L/\sqrt{4Dt}-\sqrt{\mu t}\big)\,,
\end{align}
where $ \mathrm{erfc}(z)=1-\mathrm{erf}(z)$, and
$\mathrm{erf}\, z =(2/\!\sqrt{\pi}) \int_0^z e^{-u^2} du$ is the error
function.  When the searcher is immortal (the case $\mu\!=\!0$),
Eq.~\eqref{exit-t} reduces to the more familiar expression
$F_1(t)=\mathrm{erfc}\big(L/\sqrt{4Dt}\big)$ for the probability that a
diffusing particle reaches the origin by time $t$.  The probability
${\mathcal F}_1\equiv F_1(t\!=\!\infty)$ that the target is \emph{eventually}
found is
\begin{align}
\label{exit}
{\mathcal F}_1 = \int_0^{\infty} dt\, f_1(t)= e^{-\sqrt{\mu/D}\,L}\,.
\end{align}
For an immortal searcher ${\mathcal F}_1=1$, whereas for high mortality rate
${\mathcal F}_1$ is exponentially small.

From the first-passage distribution \eqref{FPP}, the average and the variance
of the search time are
\begin{subequations}
\begin{align}
 \langle T \rangle &= \frac{\int_0^{\infty} dt\,t f_1(t)}{\int_0^{\infty} dt\, f_1(t)}
= \frac{L}{\sqrt{4 D \mu }}= \sqrt{\tau_D\tau_\mu}\,,\label{mean}\\
\Sigma^2&= \langle T^2\rangle - \langle T \rangle^2 = \frac{2 D L}{\left(4 D
    \mu\right)^{3/2}}= \frac{1}{2}\sqrt{\tau_D\tau_\mu^3}\,,
\label{var}
\end{align}
\end{subequations}
where $\tau_D\equiv L^2/4D$ is the diffusion time and $\tau_\mu\equiv 1/\mu$
is the lifetime of a mortal searcher.  Notice that low-order moments of the
search-time distribution involve non-trivial combinations of $\tau_D$ and
$\tau_\mu$.  Also notice that as $\mu\to 0$, corresponding to immortal
searchers, both $\langle T\rangle$ and $\Sigma^2$ diverge.

We now turn to the case of $N$ immortal searchers.  The probability that one
of them first reaches the target at time $t$ is~\cite{Monasterio,MR}
\begin{align}
f_N(t) &= N f_1(t) \big[1-F_1(t)\big]^{N-1}\nonumber \\
& = N \,\frac{L}{\sqrt{4\pi Dt^3}}\,\, e^{-L^2/4Dt}
  \left[\mathrm{erf}\big(L/\sqrt{4Dt}\big)\right]^{N-1}\nonumber \\
&\simeq  N \frac{L}{\sqrt{4\pi Dt^3}}\,
\left(\frac{L}{\sqrt{\pi Dt}}\right)^{N-1},  \qquad t\to\infty,
\end{align}
where the third line follows from the $z\ll 1$ asymptotic of
$\mathrm{erf}(z)$. An important feature of $f_N(t)$ is that it has the
algebraic long-time tail $f_N(t)\sim t^{-(N+2)/2}$.  Thus, as mentioned in
the introduction, the average search time is divergent for $N\leq 2$, but
finite for $N\geq 3$ \cite{Lindenberg80}.

We now determine the average time for the first out of $N\!\geq\! 3$ immortal
searchers to reach the target.  By definition
\begin{align}
\label{I1}
\langle T_N\rangle &\!=\! \int_0^\infty \!\! dt\,t \,f_N(t)\nonumber \\
&\!=\! \int_0^\infty \!\!dt\,t \,\frac{NL}{\sqrt{4\pi Dt^3}}\,\, e^{-L^2/4Dt}\!
 \left[\mathrm{erf}\big(L/\sqrt{4Dt}\big)\right]^{N-1}\!\!\!\!\!.
\end{align}
We integrate by parts and define the scaled variable $z=L/\sqrt{4Dt}$ to recast
this expression as \cite{Monasterio}
\begin{equation}
\label{t-L}
\langle T_N\rangle \!=\!\tau_D \Psi(N),\qquad \Psi(N)=2\int_0^{\infty} \frac{dz}{z^3}\,\,\,\mathrm{erf}^N z .
\end{equation}
For very large $N$, $\mathrm{erf}^N z$ effectively becomes the Heaviside step
function $\theta(z\!-\!z_0)$, where $z_0=z_0(N)\gg 1$. To determine $z_0(N)$, we
use the large-$z$ asymptotic
$\mathrm{erf}\, z \simeq 1-e^{-z^2}/(\sqrt{\pi}\,z)$, so that
$\mathrm{erf}^N z\simeq\exp [-N e^{-z^2}/(\sqrt{\pi} z)]$.  Now $z_0(N)$ is
determined, with logarithmic accuracy, from the condition
$N e^{-z^2}/z \sim 1$, or $z \,e^{z^2} \sim N$. This yields
$z_0\simeq [W(2N^2)]^{1/2}/\sqrt{2}$, where $W(\cdot)$ is the Lambert
$W$-function---the inverse of $f(W)=W e^W$ \cite{Wolfram}.  Thus
\begin{equation}\label{Psiresult}
\Psi(N) \simeq 2\int_{z_0(N)}^{\infty} \frac{dz} {z^3} = \frac{1}{z_0^2(N)} \simeq \frac{2}{W(2N^2)} .
\end{equation}
To leading order in $\ln N\gg 1$, this yields
\begin{equation}
\label{time0}
\langle T_N \rangle \simeq \frac{L^2}{4 D\,\ln N}\,.
\end{equation}
This result coincides with the asymptotic that was quoted in
Ref.~\cite{Monasterio}.  The average search time (\ref{time0}) decays only
logarithmically with $N$ but still gives to a reduction by a factor of $20$
compared with the characteristic diffusion time $\tau_D=L^2/4D$ for the
typical number of human sperm ($N=3 \times 10^8$) that attempt to fertilize
an oocyte.

For $N$ identical and mortal searchers that all start from $x=L$, the
probability that at least one of them eventually reaches the target is
\begin{equation}
\label{probN}
    p=\sum_{k=1}^{N} \binom{N}{k} {\mathcal F}_1^k (1-{\mathcal F}_1)^{N-k}=1-(1-{\mathcal F}_1)^N,
\end{equation}
with ${\mathcal F}_1$ given by Eq.~\eqref{exit}.  It is convenient to
introduce the dimensionless mortality rate $M\!=\!\sqrt{\mu\,\tau_D}$.  For
high mortality, $M\gg 1$, we approximate $p\simeq 1-\exp(-N e^{-2 M})$, which
changes rapidly from being vanishingly small for $N<N_c(M)$ to being close to
$1$ for $N>N_c(M)$, with $N_c(M)\simeq e^{2M}\gg 1$.  An important message
from this simple argument is that a huge redundancy of searchers is needed to
offset their high mortality for a search to be successful.

For mortal searchers, reaching the target is not guaranteed, and the
average time of successful search is
\begin{align}
\label{tmortal}
\langle T_N\rangle = \frac{\int_0^\infty t \,f_N(t)\, dt}{\int_0^\infty f_N(t)\, dt}
\equiv \tau_D \,\frac{\Psi_1(N,M)}{\Psi_0(N,M)}\,.
\end{align}
Here $\Psi_k(N,M) = \int_0^\infty {dz}\,{z^{-2k}} \, e^{-\Phi(N,M,z)}$ and,
by using Eqs.~\eqref{exit-t} and \eqref{I1}, $z\!=\!L/\sqrt{4Dt}$, and also
expressing all variables in scaled form, straightforward algebra gives
\begin{align}
&\Phi(N,M,z)= z^2\!+\!\frac{M^2}{z^2} \nonumber \\
&\!-\!(N\!-\!1)\!\ln\!\left[\!1\!-\!\frac{e^{2M}}{2}\mathrm{erfc}\Big(z\!+\!\frac{M}{z}\Big)
  \!-\! \frac{e^{-2M}}{2}\mathrm{erfc}\left(z\!-\!\frac{M}{z}\right)\!\right] .\nonumber
\end{align}
We are interested in the high-mortality regime, $M\gg 1$, where the
probability that a single searcher eventually reaches the target, as given in Eq.~(\ref{exit}), is
exponentially small.  The inequality $M\gg 1$
enables us to evaluate the integral in $\Psi_k$ by the standard Laplace
method~\cite{AW12}. The saddle point $z_*(N,M)$ is found by minimizing
$\Phi(N,M,z)$ with respect to $z$:
\begin{equation}\label{der10}
\frac{d\Phi(N,M,z)}{dz}\simeq 2 z-\frac{2 M^2}{z^3} -\frac{2(N\!-\!1)}{\sqrt{\pi}}\, e^{-z^2\!-\!M^2/z^2},
\end{equation}
where we again use the $z\gg 1$ asymptotic of $\mathrm{erfc}(z)$.  Since the
functions $\Psi_0$ and $\Psi_1$ include the same exponent $e^{\Phi(N,M,z)}$,
we obtain, after cancellations, $\Psi_1/\Psi_0\simeq z_*^{-2}$.

There are two distinct limiting behaviors for the saddle point $z_*$ that
depend on the interplay between $N$ and $M$.  For $N\ll \sqrt{M} e^{2M}$, $z_*$ is
determined by balancing the first two terms in Eq.~\eqref{der10}.  This
yields $z_*\simeq \sqrt{M}$.  In this regime, we obtain
\begin{equation}\label{tNa}
\langle T_N\rangle \simeq \sqrt{\tau_{\mu} \tau_D}= \frac{\tau_D}{M}=\frac{L}{\sqrt{4D\mu}}\,,
\end{equation}
which coincides with Eq.~\eqref{mean} for the average search time of a
\emph{single} searcher.  In this high mortality regime, only the fastest
searcher contributes to the average time, while the rest of the searchers
(unless their number is huge, see below) are superfluous.  

In the limit of $N\!\gg\!\sqrt{M} e^{2M}$, but with $M$ also large, the
saddle point is given by $2z - 2(N-1) e^{-z^2}/\sqrt{\pi} \simeq 0$, which
yields $z_*\simeq z_0$, the same quantity that arises for immortal searchers,
as given above Eq.~(\ref{Psiresult}).  As a result, the final expression for
$\langle T_N\rangle$ coincides with Eq.~\eqref{time0}.  Thus when the number
of searchers is extremely large, their mortality (even when relatively high)
is irrelevant in the determination of average search time.  

\begin{figure}[h]
  \includegraphics[width=3.4in,height=2.25in]{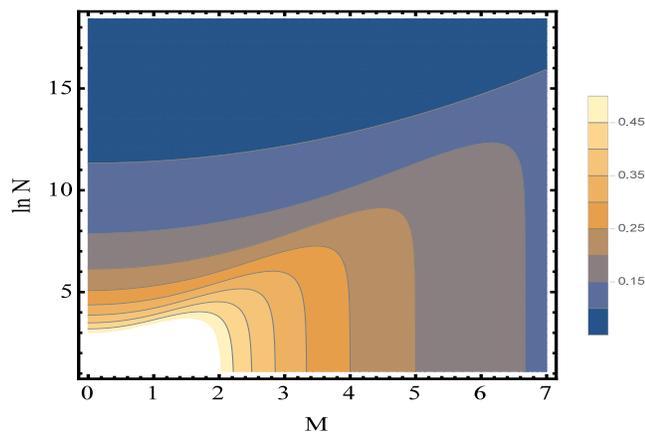}\\
  \caption{(Color online) Contour plot of $\langle T_N \rangle/\tau_D$ from
    Eq.~\eqref{tmortal} vs.\ $M$ and $\ln N$.}
   \label{contour}
\end{figure}

\begin{figure}[h]
\subfigure[]{\includegraphics[width=0.235\textwidth]{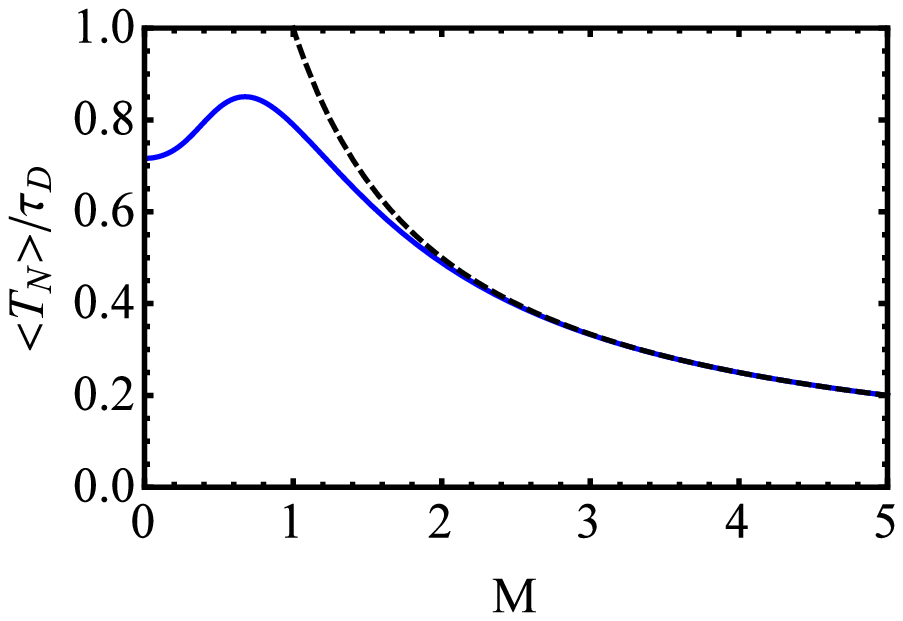}}
\subfigure[]{ \includegraphics[width=0.235\textwidth]{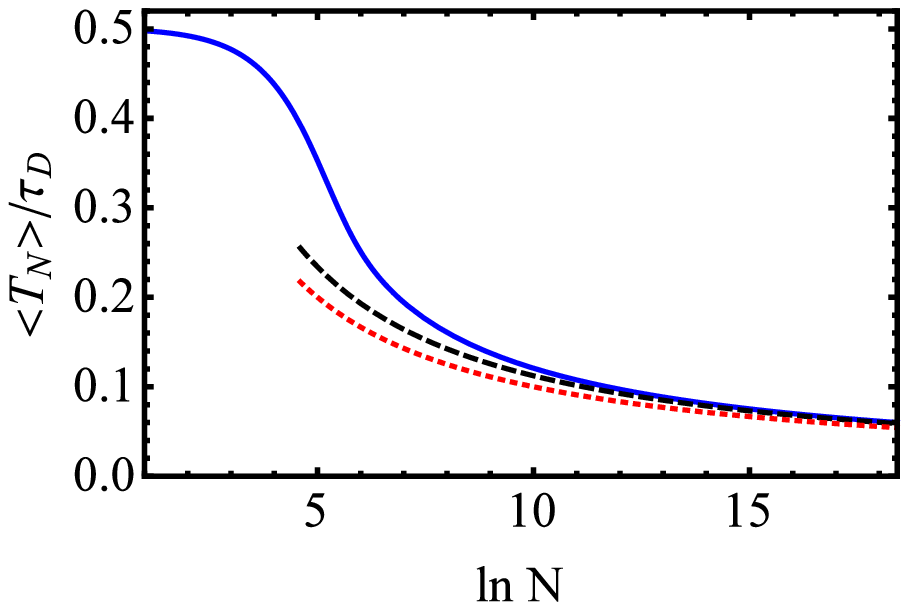}}
\caption{(Color online) (a) Non-monotonic dependence of $\langle T_N \rangle$
  from Eq.~\eqref{tmortal} on $M$ for $N=10$.  The large-$M$ asymptotic
  $\langle T_N\rangle/\tau_D\simeq 1/M$ (Eq.~\eqref{tNa}) is shown dashed.
  (b) $\langle T_N \rangle/\tau_D$ vs.\ $\ln N$ for $M=2$.  The dashed and
  dotted lines are the respective asymptotics \eqref{Psiresult} and
  \eqref{time0}.}
   \label{nonmonotonic}
\end{figure}

Figure \ref{contour} shows the dependence of $\langle T_N\rangle$ on $M$ and
$\ln N$.  For not too large $N$, $\langle T_N\rangle$ in Eq.~\eqref{tmortal}
exhibits a maximum as a function of the mortality rate, as evident from
Fig.~\ref{contour} and Fig.~\ref{nonmonotonic}(a).  This latter figure also
shows that the $M\gg1$ asymptotic $\langle T_N\rangle \simeq \tau_D/M$ is
accurate already at $M\simeq 2$.  Figure~\ref{nonmonotonic}(b) shows
$\langle T_N\rangle/\tau_D$ vs.\ $\ln N$ and the asymptotics
\eqref{Psiresult} and \eqref{time0}.

The biologically relevant situation where mortal searchers have distinct
diffusivities $D_k$, $k=1, \dots, N$~\cite{diversity} gives rise to a new
optimization criterion for a successful search.  Now the probability that a
searcher with diffusivity $D_k$ eventually reaches the target is
\begin{equation}\label{exitv0}
\mathcal{F}(D_k) = e^{-L\sqrt{\mu/D_k}} \ll 1.
\end{equation}
When $N \mathcal{F}(D_k) \ll 1$ for typical diffusivity values, we may approximate the
probability that a single searcher reaches the target as
$p\simeq \sum_{k=1}^{N} \mathcal{F}(D_k)$.
For $N\gg 1$, we may make the simplifying assumption that the diffusivity
distribution is a Gaussian that is centered about a typical value $D_0$:
$P(D)=\frac{1}{\sqrt{2\pi\sigma^2}}\,e^{-(D-D_0)^2/2\sigma^2}$,
where $\sigma \ll D_0$.  Then the summation for $p$ may be replaced by the
integral
\begin{align}
\label{probint}
p&\simeq \int_0^{\infty} dD\, \mathcal{F}(D) P(D) \nonumber \\
&\simeq \frac{1}{\sqrt{2\pi\sigma^2}}\,
\int_0^{\infty} dD\, e^{- L\sqrt{\mu/D}-(D-D_0)^2/2 \sigma^2}.
\end{align}

We again evaluate the integral by the Laplace method~\cite{AW12}.  By
straightforward rescaling, we redefine the above integral as
\begin{equation}\label{probint2}
p=\frac{D_0}{\sqrt{2\pi\sigma^2}} \int_0^{\infty} dz\,e^{-L\sqrt{\frac{\mu}{D_0}}\,\Phi(z)},
\end{equation}
where $\Phi(z)=z^{-1/2}+A(z-1)^2$, with
\begin{equation*}
A=\frac{D_0^2}{2\sigma^2} \times \frac{\sqrt{D_0}}{L\sqrt{\mu}}\quad\mathrm{and}\quad
z=\frac{D}{D_0}\,.
\end{equation*}
We find the saddle point $z_*$ from the equation
$\Phi^{\prime}(z)=2A (z-1)-(1/2) z^{-3/2}=0$. The exact solution of this
equation is cumbersome, and we confine ourselves to the asymptotics in the
limits $A\gg 1$ and $A\ll 1$.  In the low mortality limit where $A\gg 1$ (but
still $L\sqrt{\mu/D_0}\gg 1$), $z_*$ is very close to 1, and we find
$z_*\simeq 1+(4A)^{-1}$.  Applying the Laplace method, the
arrival probability is
\begin{equation}\label{largeA}
p\simeq \exp\left(-\frac{L\sqrt{\mu}}{\sqrt{D_0}}\,\right), \qquad A\gg 1.
\end{equation}
In this case, the dominant contribution to $p$ comes from ``typical''
searchers---those whose diffusivity is close to $D_0$.  This behavior is to
be expected, as in the low-mortality limit the Gaussian in the integrand of
Eq.~\eqref{probint} effectively acts as a delta-function peak centered
at $D_0$.

The high-mortality limit, $A\ll 1$, is more interesting.  Here $z_*\gg 1$,
and we can replace $z-1$ with $z$ in $\Phi(z)$ to arrive at
$z_*\simeq (4A)^{-2/5}$.  In this case, a small number of highly active
searchers with $D\gg D_0$ give the dominant contribution to the probability
that the target is found.  Performing the Gaussian integral, we obtain
\begin{equation}\label{smallA}
  p \simeq\sqrt{\frac{2}{5}} \exp\left[-\Big(2^{1/5}\!+\!2^{-9/5}\Big)\,
  \Big(\frac{\mu L^2}{\sigma}\Big)^{2/5}\right], \qquad A\ll 1.
\end{equation}
Surprisingly, this result for the arrival probability is \emph{independent}
of the average diffusivity $D_0$.  Numerical integration of \eqref{probint2}
shows that the arrival probability decays monotonically with the basic
dimensionless parameter $A$, and the asymptotic forms \eqref{largeA} and
\eqref{smallA} match the full solution in the respective limits of $A\ll 1$
and $A\gg 1$.

There are two additional attributes that naturally arise in the phenomenon of
oocyte fertilization by sperm that can be accounted for within the present
framework.  One is the possibility of a diversity in searcher lifetimes.  If
the mortality rate $\mu$ is normally distributed, we find that the search
dynamics is only trivially affected~\cite{mortalitydist}.  In this case, the
main contribution to the probability that an ensemble of searchers reaches
the target comes from typical searchers.

The role of chemotaxis, which is known to be a dominant effect as sperm
approach the fertilization site~\cite{Eisenbach}, has more interesting
consequences.  A simplistic way to model chemotaxis is to include a constant
drift velocity $v$ in the diffusion-decay model.  The probability evolution
of a single searcher is now governed by the equation of motion
$\partial_t \rho - v\partial_x \rho =-\mu \rho +D\partial_x^2 \rho$, where
$v>0$.  When a single searcher starts at $x=L$ and seeks a target that is at
the origin, the solution to this equation, subject to the boundary conditions
$\rho(x\!=\!0,t)=\rho(x\!=\!\infty, t)=0$ and the initial condition
$\rho(x,t\!=\!0)=\delta (x-L)$, can again be found by the image
method~\cite{R01}.  The result is
\begin{align}
\label{solv}
    \rho(x,t)&=\frac{1}{\sqrt{4\pi Dt}}\,e^{-\mu t}
   \left[e^{-(x+v t-L)^2/4Dt}\right.\nonumber \\
&\qquad\qquad\qquad\left. -e^{vL/D}e^{-(x+v t +L)^2/4Dt}\right].
\end{align}
From this expression, the probability density that the target is reached at
time $t$ is
\begin{equation}
\label{FPP-v}
  f_1(t)=\frac{L}{\sqrt{4\pi D t^3}}\,\,  e^{-\mu t -{(L-vt)^2}/{4 D t}}\,.
\end{equation}
Consequently, the probability that the target is eventually reached is
\begin{align}
\label{exitv}
{\mathcal F}_1 = \int_0^{\infty} dt\, f_1(t)= e^{-\left(\sqrt{4 D \mu +v^2}-v\right)L/2 D}\,.
\end{align}
From Eq.~\eqref{FPP-v}, the average search time and its variance are
\begin{align}
\label{mean-v}
  \langle T \rangle =  \frac{L}{\sqrt{4 D \mu +v^2}}\,,\qquad
  \Sigma^2=  \frac{2 D L}{\left(4 D \mu +v^2\right)^{3/2}}\,.
\end{align}
Interestingly, the variance is non-monotonic in $D$, and the largest
fluctuations in the search time occur when $D=v^2/(2\mu)$.  Using our general
approach, we can extend this chemotactic search process to the case of many
searchers and determine the behavior of the search time on the $N$, $M$, and
the rescaled drift velocity, the P\'{e}clet number~\cite{R01}.  It is also
possible to investigate this search process in higher dimensions and in more
realistic geometries, with the goal of providing a realistic but still
tractable model for fertilization.

To summarize, we elucidated the competing roles of mortality, redundancy, and
diversity on a search process that represents a caricature for the
fertilization of an oocyte by sperm.  To optimize this search, one strategy
is to have a sophisticated search algorithm.  However, nature often seems to
prefer the brute-force approach of dispatching many almost-identical
searchers that follow a simple search algorithm.  The effectiveness of this
redundancy is counterbalanced by the mortality of the searchers, and it is
miraculous that the correct (and a very small) number of sperm actually reach
the oocyte in human fertilization.  While we do not offer insight into why
this miracle occurs, we quantified the dynamics of this search process as a
function of the number of searchers and their mortality rate.  We also found
that searcher diversity can compete with mortality so that only the most
active searchers are successful.

We thank David Holcman and Nataly Meerson for a discussion of the sperm
search problem.  Financial support for this research was provided in part by
grant No.\ 2012145 from the United States-Israel Binational Science
Foundation (BSF) (SR and BM) and Grant No.\ DMR-1205797 from the NSF (SR).


\bigskip\bigskip

\begin{thebibliography}{99}

\bibitem{BLMV11} For a recent review from the physics perspective, see, e.g.,
  O. B\'enichou, C. Loverdo, M. Moreau, and R. Voituriez, Rev.\ Mod.\ Phys.\
  {\bf 83}, 81 (2011).

\bibitem{BWV81} O. G. Berg, R. B. Winter and P. H. Von Hippel, Biochem.\
  {\bf 20}, 6929 (1981).

\bibitem{V07} P. H. Von Hippel, Ann.\ Rev.\ Biophys.\ Biomol. Struct.\ {\bf
    36}, 79 (2007).

\bibitem{M08} L. Mirny, Nature Physics {\bf 4}, 93 (2008).

\bibitem{C76} E. L. Charnov, Theor.\ Popul.\ Biol.\ {\bf 9}, 129 (1976).

\bibitem{B91} W. J. Bell, \textit{Searching Behaviour: the Behavioural
    Ecology of Finding Resources} (Chapman and Hall, London, 1991).

\bibitem{OBE90} W. J. O'Brien, H. I. Browman, and B. I. Evans, Am.\ Sci.\
  {\bf 78}, 152 (1990).

\bibitem{VBH99} G. M. Viswanathan, S. V. Buldyrev, S. Havlin, M. G. E. Da
  Luz, E. P. Raposo, and H. E. Stanley, Nature {\bf 401}, 911 (1999).

\bibitem{BLMV06} O. B\'enichou, C. Loverdo, M. Moreau, and R. Voituriez,
  Phys.\ Rev.\ E {\bf 74}, 020102 (2006).

\bibitem{S17} M. von Smoluchowski, Zeitschrift f\"{u}r physikalische
  Chemie, Stochiometrie und Verwandtschaftslehre {\bf 92}, 129 (1917).

\bibitem{HTB90} P. H\"{a}nggi, P. Talkner, and M. Borkovec, Rev.\ Mod.\
  Phys. {\bf 62}, 251 (1990).

\bibitem{RS71} H. R. Richardson and L. D. Stone, Naval Research Logistics
  Quarterly {\bf 18}, 141 (1971).

\bibitem{FS01} J. R. Frost and L. D. Stone,
  \url{http.//www.rdc.uscg.gov/reports/2001/cgd1501dpexsum.pdf}.

\bibitem{S09} M. F. Shlesinger, J. Phys.\ A: Math.\ and Theor.\ {\bf 42}, 2009.

\bibitem{BNHW87} R. F. Bonner, R. Nossal, S. Havlin, and G. H. Weiss,
J. Opt.\ Soc.\ Am.\ {\bf A4}, 423 (1987).

\bibitem{YAL13} S. B. Yuste, E. Abad, and K. Lindenberg, Phys.\ Rev.\ Lett.\
  {\bf 110}, 220603 (2013).

\bibitem{AYL13} E. Abad, S. B. Yuste, and K. Lindenberg, Phys.\ Rev.\ E {\bf
    88}, 062110 (2013).

\bibitem{M15} B. Meerson,  arXiv:1502.02813.

\bibitem{BR14} O. B\'enichou and S. Redner, Phys.\ Rev.\ Lett.\ {\bf 113},
  238101 (2014).

\bibitem{Eisenbach} M. Eisenbach and L. C. Giojalas, Nature Reviews Molecular
  and Cell Biology \textbf{7}, 276 (2006).

\bibitem{Holcman} K. Reynaud, Z. Schuss, N. Rouach, and D. Holcman, arXiv 1409.7941.

\bibitem{ZKB83} G. Zumofen, J. Klafter, and A. Blumen, J. Chem.\ Phys.\ {\bf
    79}, 5131 (1983).

\bibitem{RK84} S. Redner and K. Kang, J. Phys.\ A {\bf 17}, L451 (1984).

\bibitem{BZK84} A. Blumen, G. Zumofen, and J. Klafter, Phys.\ Rev.\ B
{\bf 30}, 5379(R) (1984).

\bibitem{Blythe} R. A. Blythe and A. J. Bray, Phys.\ Rev.\ E \textbf{67},
  041101 (2003).

\bibitem{BMS13} A. J. Bray, S. N. Majumdar, and G. Schehr, Adv.\ Phys.\
{\bf 62}, 225 (2013).

\bibitem{RM} S. Redner and B. Meerson, J. Stat.\ Mech.\ (2014) P06019.

\bibitem{MVK} B. Meerson, A. Vilenkin, and P. L. Krapivsky, Phys.\ Rev.\ E
  \textbf{90}, 022120 (2014).

\bibitem{Lindenberg80} K. Lindenberg, V. Seshadri, K. E. Shuler, and
  G. H. Weiss, J. Stat.\ Phys.\ \textbf{23}, 11 (1980).

\bibitem{RK99} S. Redner and P. L. Krapivsky, Am.\ J. Phys.\ \textbf{67},
  1277 (1999).

\bibitem{KMR} P. L. Krapivsky, S. N. Majumdar, and A. Rosso, J. Phys.\ A:
  Math.\ Theor.\ \textbf{43}, 315001 (2010).

\bibitem{Monasterio} C. M.-Monasterio, G. Oshanin, and G. Schehr, J. Stat.\
  Mech.\ (2011) P06022.

\bibitem{MR} B. Meerson and S. Redner, J. Stat.\ Mech.\ (2014) P08008.

\bibitem{R01} S.~Redner, {\it A Guide to First-Passage Processes} (Cambridge
  University Press, Cambridge, 2001).

\bibitem{Wolfram} Wolfram Research, Inc., Mathematica, Version 10.0, Champaign, IL (2014).

\bibitem{AW12} G. B. Arfken and H. J. Weber, {\it Mathematical Methods for
    Physicists}, $7^{\rm th}$ ed.\ (Academic Press, New York, 2012).

\bibitem{diversity} See e.g., C. \'Alvarez, J. A. Castilla, L. Mart\'inez,
  J. P. Ram\'irez, F. Vergara, and J. J. Gaforio, Human Reproduction
  \textbf{18}, 2082 (2003); \textit{Sperm Biology: An Evolutionary
    Perspective}, Eds.\ T. R. Birkhead, D. J. Hosken, and S. S. Pitnick
  (Academic Press, Burlington MA, 2009).

\bibitem{mortalitydist} The derivation is similar to that for a distribution
  of diffusivities, as presented in the material associated with Eq.~\eqref{probint}.


\end{thebibliography}
\end{document}